\documentclass[12pt,preprint]{aastex}
\usepackage{psfig}
\usepackage{graphicx}

\shorttitle{Modeling a loop with MHD nanoflares}
\shortauthors{Reale et al.}

\begin{document}

\def\ntilde{\hbox{\rm n}}
\def\vv{\hbox{\bf v}}
\def\gvec{\hbox{\bf g }}
\def\rvec{\hbox{\bf r }}
\def\svec{\hbox{\bf s }}
\def\vvec{\hbox{\bf v }}
\def\solphys{\hbox{Sol. Phys.}}
\def\aaps{\hbox{A\&AS}}
\def\aap{\hbox{A\&A}}
\def\apj{\hbox{ApJ}}
\def\apjl{\hbox{ApJL}}
\def\nat{\hbox{Nature}}

\title{Modeling a coronal loop heated by MHD-turbulence nanoflares}

\author{F. Reale\altaffilmark{1}, G. Nigro\altaffilmark{2}, F.
Malara\altaffilmark{2}, G. Peres\altaffilmark{1}, P.
Veltri\altaffilmark{2}}

\altaffiltext{1}{Dipartimento di Scienze Fisiche \& Astronomiche, Sezione di
Astronomia, Universit\`a di Palermo, Piazza del Parlamento 1,
I-90134 Palermo, Italy}

\altaffiltext{2}{Dipartimento di Fisica, Universit\`a della Calabria,
I-87030 Arcavacata di Rende, Italy}


\begin{abstract}
We model the hydrodynamic evolution of the plasma confined in a coronal
loop, 30000 km long, subject to the heating of nanoflares due to
intermittent magnetic dissipative events in the MHD turbulence produced
by loop footpoint motions. We use the time-dependent distribution of
energy dissipation along the loop obtained from a hybrid shell model,
occurring for a magnetic field of about 10 G in corona; the relevant
heating per unit volume along the loop is used in the Palermo-Harvard
loop plasma hydrodynamic model. We describe the results focussing on
the effects produced by the most intense heat pulses, which lead to loop
temperatures between 1 and 1.5 MK.
\end{abstract}

\maketitle

\section{Introduction}

Nanoflares (Parker 1988) are among the best candidates to explain the
heating of the solar corona, and, in particular, of the coronal loops
(e.g. Peres et al. 1993, Cargill 1993, Kopp \& Poletto 1993, Shimizu
1995, Judge et al. 1998, Mitra-Kraev \& Benz 2001, Katsukawa \& Tsuneta
2001, Warren et al. 2002, 2003, Spadaro et al. 2003, Cargill \&
Klimchuk 1997, 2004, M\"uller et al. 2004, Testa et al. 2004).

Although the evidence of nanoflares appears to be well established, it is
still unclear whether, and to what extent, they really can provide enough
energy to heat the whole corona (e.g. Aschwanden 1999).  More recently,
models of nanoflares with a prescribed random time distribution of the
pulses deposited at the footpoints of multi-stranded loops have been
proposed (Warren et al. 2002, Warren et al. 2003), and have been shown
to describe several observed features.

According to some models, nanoflares are the result of dissipation in
an MHD turbulence, generated inside closed magnetic structures in the
corona, and due to nonlinear interactions among fluctuations generated
by photospheric motions. Possible evidence of turbulent motions has been
detected from line broadenings in coronal loops (Saba \& Strong 1991).
Most of these models include direct numerical solution of MHD equations in
two or three dimensions (Einaudi et al., 1996; Hendrix \& Van Hoven, 1996;
Dmitruk \& G\'omez, 1997; Dmitruk et al., 1998; Dmitruk \& G\'omez, 1999;
Buchlin et al., 2003) using relatively low Reynolds/Lundquist numbers.
Recently Nigro et al. 2004 (hereafter NMCV04) have related coronal
nanoflares to intermittent dissipative events in the MHD turbulence
produced in a coronal magnetic structure by footpoint motions. The
injected energy is stored in the loop up to significant levels in the
form of magnetic and velocity fluctuations and released intermittently
through nonlinear interactions which process these fluctuations and
generate cascades toward smaller scales where energy is dissipated. The
derived probability distribution functions of the peak maximum power,
peak duration time, energy dissipated in a burst and waiting time between
bursts are in good agreement with those obtained from the analysis of
coronal impulsive events (Datlowe et al.  1974, Lin et al. 1984, Dennis
1985, Crosby et al. 1993, Shimizu \& Tsuneta 1997, Krucker \& Benz 1998,
Boffetta et al. 1999, Parnell \& Jupp 2000, Aschwanden et al. 2000a, b).
This heating model does not need any {\it ad hoc} hypothesis, once the
loop length and the characteristic Alfven speed, i.e. the strength of the
ambient magnetic field (if the density does not change much), are fixed.

In the present work we model the plasma confined in a coronal loop heated
according to the events dissipation rate and distribution described
in NMCV04. We will compute the evolution of the distributions of the
density, temperature and velocity of the loop plasma by means of the
time-dependent thermo-hydrodynamic Palermo-Harvard (Peres et al. 1982,
Betta et al.  1997) loop model assuming the output of the hybrid shell
model illustrated in NMCV04 as the basis of the heating function.

In Section~\ref{sec:model} we describe the set up of the loop model
with the MHD-turbulence dissipation rate as input heating; in
Sec.~\ref{sec:results} we show relevant results and discuss them in
Sec.~\ref{sec:discuss}.

\section{The loop model}
\label{sec:model}

Our purpose here is to model the evolution of the plasma confined in
a coronal loop under the effect of the energy dissipation predicted
in NMCV04.  According to their settings, we model a magnetic loop, with
a total length of 30,000 km. The plasma is described as a compressible
fluid moving and transporting energy only along the magnetic field lines,
i.e. along the loop itself. Thus, the magnetic field has only the role of
confining the plasma.  The loop model assumes constant loop cross-section.

We use the Palermo-Harvard code (Peres et al. 1982, Betta et al. 1997),
a 1-D hydrodynamic code that consistently solves the
time-dependent density, momentum and energy equations for the plasma
confined by the magnetic field:

\begin{equation}
 \label{eqm:density}
    \frac{dn}{dt}= -n \frac{\partial v}{\partial s},
\end{equation}
\begin{equation}
 \label{eqm:momentum}
    n m_{\rm H} \frac{dv}{dt} = - \frac{\partial p}{\partial s}
    + n m_{\rm H} g + \frac{\partial}{\partial s}
    (\mu \frac{\partial v}{\partial s}),
\end{equation}
\begin{equation}
 \label{eqm:energy}
    \frac{d\epsilon}{dt} +(p+\epsilon) \frac{\partial v}{\partial s} =
    H - n^2 \beta P(T) + \mu (\frac{\partial v}{\partial s})^2
    + \frac{\partial}{\partial s}(\kappa T^{5/2} \frac{\partial T}{\partial
s}),
\end{equation}
with $p$ and $\epsilon$ defined by:
\begin{equation}
 \label{eqm:pressener}
    p = (1+\beta) n K_{\rm B} T \hspace{1cm}
            \epsilon =\frac{3}{2} p + n \beta \chi,
\end{equation}
where $n$ is the hydrogen number density, $s$ the spatial coordinate
along the loop, $v$ the plasma velocity, $m_{\rm H}$ the mass of
hydrogen atom, $\mu$ the effective plasma viscosity, $P(T)$ the
radiative losses function per unit emission measure, $\beta$ the
fractional ionization, i.e.\ $n_{\rm e}/n_{\rm H}$, $\kappa$ the
thermal conductivity Spitzer (1962), $K_{\rm B}$ the Boltzmann
constant, and $\chi$ the hydrogen ionization potential.  $H(s,t)$ is
a function of both space and time which describes the heat input in the
loop.  This function will be described in detail in
Sec.~\ref{sec:heating}.  The numerical code uses an adaptive spatial grid
to follow adequately the evolving profiles of the physical quantities,
which can vary dramatically in the transition region and under the effect
of the evolution.  The loop is not symmetric, the apex is at half the
numerical grid and there is a chromosphere on each side.  The boundary
conditions at the loop footpoints are the same as in Reale et al. (2000).

\subsection{The heating function}
\label{sec:heating}

The original version of the Palermo-Harvard hydrodynamic code includes
a space- and time-dependent heating function, which describes the input
of external energy triggering transient events (Peres et al. 1987).
Several formulations are possible and the code can be easily adapted.
For the present work, the heating function is given by the output
dissipation rate of NMCV04 (in the form of a numerical table).

\begin{figure*}
\centerline{\psfig{figure=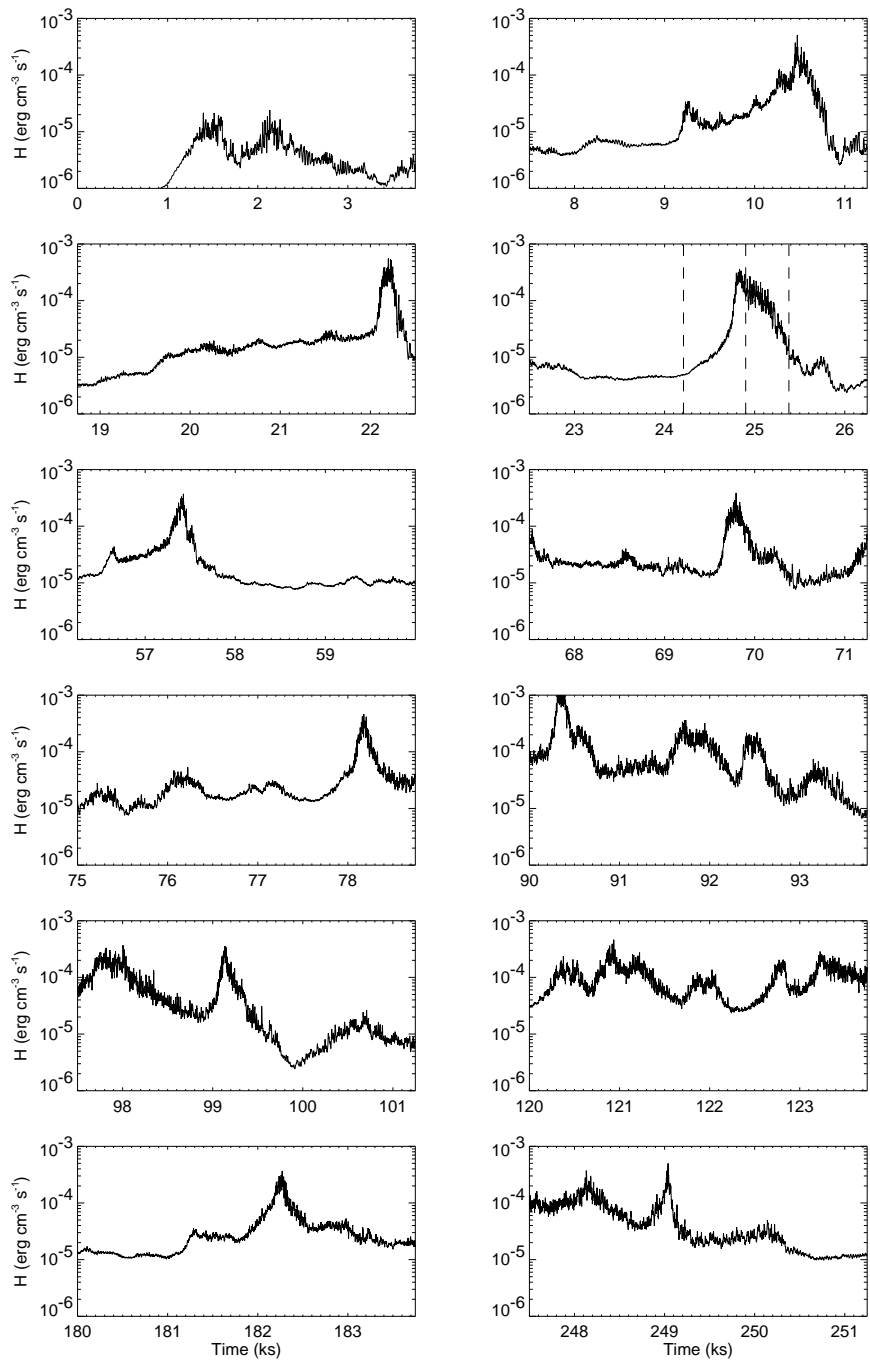,width=12cm}}
\caption[]{Evolution of the average heating rate per unit volume
released in the loop. The vertical dashed lines mark the times 
illustrated in detail in Fig.~\ref{fig:hh2d}.
\label{fig:hhlc}}
\end{figure*}

The model developed in NMCV04 has been derived within the framework of the
Reduced Magnetohydrodynamics (RMHD) (Strauss 1976, Zank \& Matthaeus,
1992), with the assumptions that: (i) the plasma is permeated by a
strong uniform magnetic field ${\bf B}_0$ in the longitudinal direction;
(ii) there is low thermal to magnetic pressure ratio $\beta_P=8\pi p/B^2
\ll 1$; (iii) the longitudinal scale $\l_{||}$ of  transverse velocity
$v_\perp$ and magnetic field $B_\perp$ fluctuations is much larger than
the transverse scale $l_\perp$; indeed, the MHD turbulence is anisotropic
(e.g., Carbone \& Veltri, 1990), the energy cascade being more efficient
perpendicularly to ${\bf B}_0$.  (iv) Small amplitude perturbations
$B_\perp /B_0 = v_\perp / c_{A0} < l_\perp / l_{||} \ll 1$, where $c_{A0}$
is the background Alfv\'en velocity, commonly assumed of the order of
$c_{A0} \sim 10^8$ cm s$^{-1}$, while the fluctuating velocity can
be estimated using nonthermal broadening of coronal spectral lines:
$v_\perp \sim 3 \times 10^6 - 1.5 \times 10^7$ cm s$^{-1}$. Under the
above assumptions the set of the RMHD equations can be derived; they
describe the evolution of magnetic and velocity fluctuations in terms of
two distinct effects: (a) wave propagation in the longitudinal direction,
at the Alfv\'en velocity; (b) nonlinear couplings, which generate a
turbulent cascade perpendicularly to ${\bf B}_0$. The model proposed by
NMCV04 (hybrid shell model) includes both these dynamical mechanisms,
but nonlinear effects are described in a simplified way by using a shell
technique (Boffetta et al., 1999): a Fourier expansion is carried out
in the perpendicular directions and the resulting spectral space is
divided into concentric shells of exponentially increasing radius. In
each shell velocity and magnetic field fluctuations are represented
by complex scalar quantities.  Nonlinear effects are reproduced by
quadratic terms representing the interactions between nearest and next
nearest neighbor shells; the coefficients are chosen so as to conserve 2D
quadratic invariants: total energy, cross helicity and squared magnetic
potential. The equation of the hybrid shell model is written as:

\begin{eqnarray}\label{shell}
&&
\left( {\partial \over \partial t} -\sigma {\partial \over \partial s}
\right)
Z^\sigma_n(x,t) = - { \chi \: k_n^2} Z^\sigma_n(s,t) +
\\
\nonumber
&&
{\rm i}k_n\Bigg( \frac{13}{24}Z^\sigma_{n+2} Z_{n+1}^{-\sigma} +
\frac{11}{24}Z^{-\sigma}_{n+2} Z^\sigma_{n+1}-
\frac{19}{48}Z^\sigma_{n+1}Z^{-\sigma}_{n-1}-
\\
&&
\frac{11}{48}Z^{-\sigma}_{n+1}Z^\sigma_{n-1}+
\frac{19}{96}Z_{n-1}^\sigma Z_{n-2}^{-\sigma}-
\frac{13}{96}Z_{n-1}^{-\sigma} Z^\sigma_{n-2})\Bigg)^*
\nonumber
\end{eqnarray}

\noindent
where $Z_n^{\sigma}(s,t) =v_{n\perp}(s,t)+\sigma b_{n\perp}(s,t)$ (with
$n=0,1,...,n_{max}$ and $\sigma =\pm 1$) are the Els\"asser variables;
$k_n = k_0 2^n$ the transverse wavenumber, with $k_0=2\pi (L/L_\perp)$;
$\chi = \lambda /(c_{A0}L)$, where the magnetic diffusivity $\lambda$
has been assumed equal to the transverse kinematic viscosity; the
asterisk means complex conjugate. Lengths are
normalized to the loop length $L$, and time to the Alfv\'en
transit time $t_A=L/c_{A0}$; the velocity $v_{n\perp}$ and magnetic
field $b_{n\perp}$ fluctuations are normalized to $c_{A0}$ and $B_0$,
respectively.

The shell technique allows us to describe the turbulence at high
Reynolds/Lundquist numbers with a relatively small number of degrees of
freedom.  In particular, we used a number of shells $n_{max}=11$, with a
very small dissipation coefficient $\chi =10^{-7}$. Since the
longitudinal spatial dependence is retained, the hybrid shell model can
describe effects of longitudinal resonance. Moreover, it is possible to
implement boundary conditions to describe the effects of transverse
motions at the loop bases. In particular, the system is excited through
the boundary at $s=0$, by imposing a given velocity perturbation at
large transverse scales, simulating photospheric motions. This
boundary perturbation amounts to $\sim 10^5$ cm s$^{-1}$, is
gaussian distributed and has a correlation time $t_c = 300$ s. At
the other boundary $s=1$ total reflection conditions are imposed. The
equations (\ref{shell}) are numerically solved using second order
finite difference schemes, both in space and in time.

During the evolution fluctuating energy enters or exits 
the driven boundary, so the total energy content in the loop fluctuates
erratically in time. At the same time nonlinear effects transfer energy
to smaller transverse scales, thus building a turbulence spectrum.
Dissipation takes place mainly at the smallest scales. Occasionally,
the velocity imposed at the lower boundary drives the loop near to one
longitudinal resonance: then, the velocity fluctuations increase at the
driven large scale shells, enhancing the energy cascade process towards
small dissipative scales. This process results in a spike of dissipated
energy, converted to heat. The dissipated power at time
$t$ and position $s$ along the loop is calculated as:

\begin{equation}\label{Wdiss}
H(s,t) = \frac{\chi}{2} \sum_{\sigma ,n} k_n^2 |Z_n^{\sigma}(s,t)|^2 
\end{equation}

\noindent
and is the heating input in the loop plasma model (Eq.~\ref{eqm:energy}).
The hybrid shell model yields the energy distribution along the loop 
integrated in the transverse direction, and provides therefore the heat
input for the one-dimensional loop model. 
The power in the whole loop is:

\begin{equation}\label{Wdiss}
W(t) = \int_0^1 H(s,t) ds
\end{equation}

\noindent
The profile of $W(t)$ contains a sequence of spikes of different
amplitudes and durations.  The space and time profile of the heating
function results from the interplay between the external driver
(photospheric motions), the loop resonance and the nonlinear turbulent
cascade.

The heat spatial distribution is sampled every 0.1 Alfven time.  For an
Alfven speed of $2 \times 10^8$ cm/s, one Alfven transit time is 15 s
(NMCV04).  The numerical table yields the heat distribution per unit
time and volume along the loop (sampled every 37.5 km) and span a
total time of 307.5 ks, i.e. 3.56 days.  We assume a circular cross
section and an aspect ratio d/L=0.2, where d is the cross-section
diameter; the cross-section area is $A = 2.83 \times 10^{17}$ cm$^2$.
Fig.~\ref{fig:hhlc} shows a few selected segments of the evolution of
the average loop heating rate $W(t)/(A~L)$; they are essentially zooms
of the dissipation power shown in Fig.~1 in NMCV04.  The heating per
unit volume is negligible in the first 1000 s.  After this (relatively
short) transient, the heating is steadily above $10^{-6}$ erg cm$^{-3}$
s$^{-1}$. The evolution of the average heating rate is highly irregular,
with sharp pulses whose duration spans all time scales from few seconds
to a few ks.  Some pulses resemble flares.  Also the pulses intensity is
highly irregular. Most of them are entirely below $10^{-4}$ erg cm$^{-3}$
s$^{-1}$.  A few of them are higher (although mostly below $10^{-3}$
erg cm$^{-3}$ s$^{-1}$); in fact, eleven heating pulses reach values
well above $3 \times 10^{-4}$ erg cm$^{-3}$ s$^{-1}$ and occur around
10.5, 22, 25, 57.5, 69.5, 78, 90, 99, 121, 182, 249 ks, as shown in
Fig.~\ref{fig:hhlc}.  The most intense pulse is the seventh one (90 ks)
and is higher than $10^{-3}$ erg cm$^{-3}$ s$^{-1}$.  The high pulses are
noticeably less frequent in the second half of the heating time interval:
nine of them occur in the first 150 ks.  Most of these pulses last $\sim
0.3-1$ ks and are rather peaked.

The heating rate per unit volume averaged over the whole heating
duration is $\approx 3 \times 10^{-5}$ erg cm$^{-3}$ s$^{-1}$.
According to the loop scaling laws (Rosner et al. 1978), for the
prescribed length this is the heating rate (per unit volume) of a loop
at an equilibrium base pressure of $\approx 0.025$ dyne cm$^{-2}$ and a
maximum temperature of $\approx 5 \times 10^5$ K.

\begin{figure}
\centerline{\psfig{figure=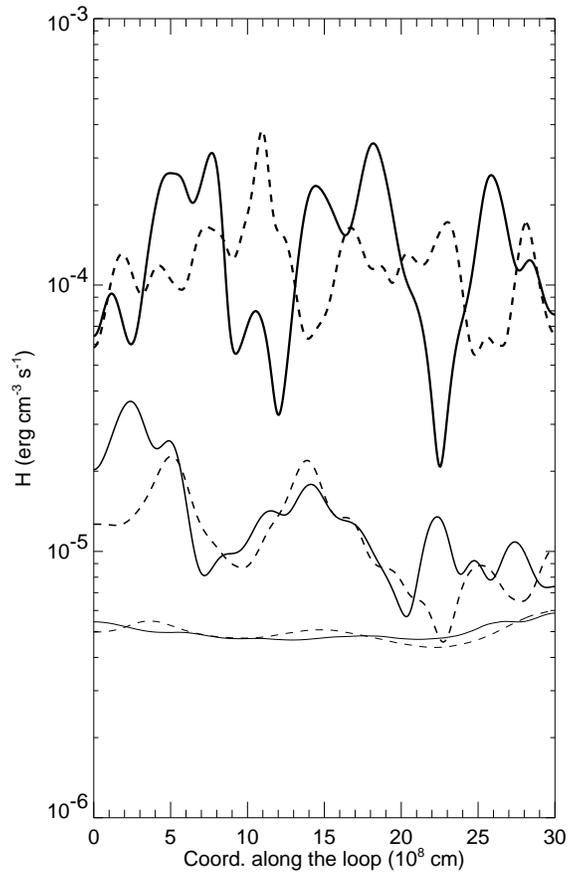,width=8cm}}
\caption[]{Spatial distributions of the heating rate per unit volume along the
loop sampled during the fourth segment of
Fig.~\ref{fig:hhlc}, at the times marked in
the fourth panel of
Fig.~\ref{fig:hhlc}. The dashed lines are the distributions after 1.5 s
from the closest solid line.
\label{fig:hh2d}}
\end{figure}

Fig.~\ref{fig:hh2d} shows distributions of the heating rate per
unit volume along the loop sampled during the fourth segment in
Fig.~\ref{fig:hhlc} (from 22.5 to 27 ks, hereafter {\it segment Ref1}).
For each time, a couple of distributions are shown, one at 1.5 s from the
other.  The heating distribution is quite uniform for low heating. During
the high intensity phase of the heating, the distribution becomes less
uniform, with large peaks propagating back and forth along the loop and
extending over $\sim 1/5$ of the loop.

\subsection{The initial conditions}

Since our scope is to investigate the structure, stability, and observable
properties of the simulated loop both in time and on the average, the
initial conditions ought to be moderately important: we should start
with an initially cool and empty loop, thereafter entirely governed by
the new time-dependent heating. For technical reasons, our choice has
been to set up this condition by letting an initially hotter loop relax
to a much cooler condition. The initial loop is obtained from the model
of Serio et al. (1981) with a uniform steady heating and a base pressure
0.03 dyne cm$^{-2}$, corresponding to a loop maximum temperature
of $\approx 5 \times 10^5$ K, i.e. the expected average condition of
the nano-flare heated loop.  In order to let this loop relax, we made
a preliminary time-dependent simulation assuming zero coronal heating
in the loop (but keeping the chromospheric heating on, to have stable
footpoints). The simulation followed the loop evolution for 2000 s,
i.e. approximately 2.5 loop thermal decay times (Serio et al. 1991). At
the end of the simulation, the loop maximum temperature decreased to
$\sim 60,000$ K, and the pressure to $\approx 1.5 \times 10^{-4}$
dyne cm$^{-2}$. A residual velocity field was present in the loop,
with speeds not larger than 6 km/s, an amply subsonic (Mach 0.2) value.
We took this final status as the initial condition for the simulations
with the nanoflare heating.

\section{Results}
\label{sec:results}

Our main purpose here is to explore how the dissipation rate described
in NMCV04 can bring a loop to coronal conditions and maintain it.
In this perspective we will describe in detail the solution obtained
in a segment containing a heat pulse of medium intensity, specifically
the fourth segment (named {\it Ref1}, between 22.5 and 26.3 ks) in
Fig.~\ref{fig:hhlc}. We will also discuss the segment including the
highest heat pulse, i.e.  the eighth segment (which we will call {\it
RefH}).  The solutions in the other segments do not differ much from
those that we are going to illustrate.

\subsection{Medium pulse}

\begin{figure*}
\centerline{\psfig{figure=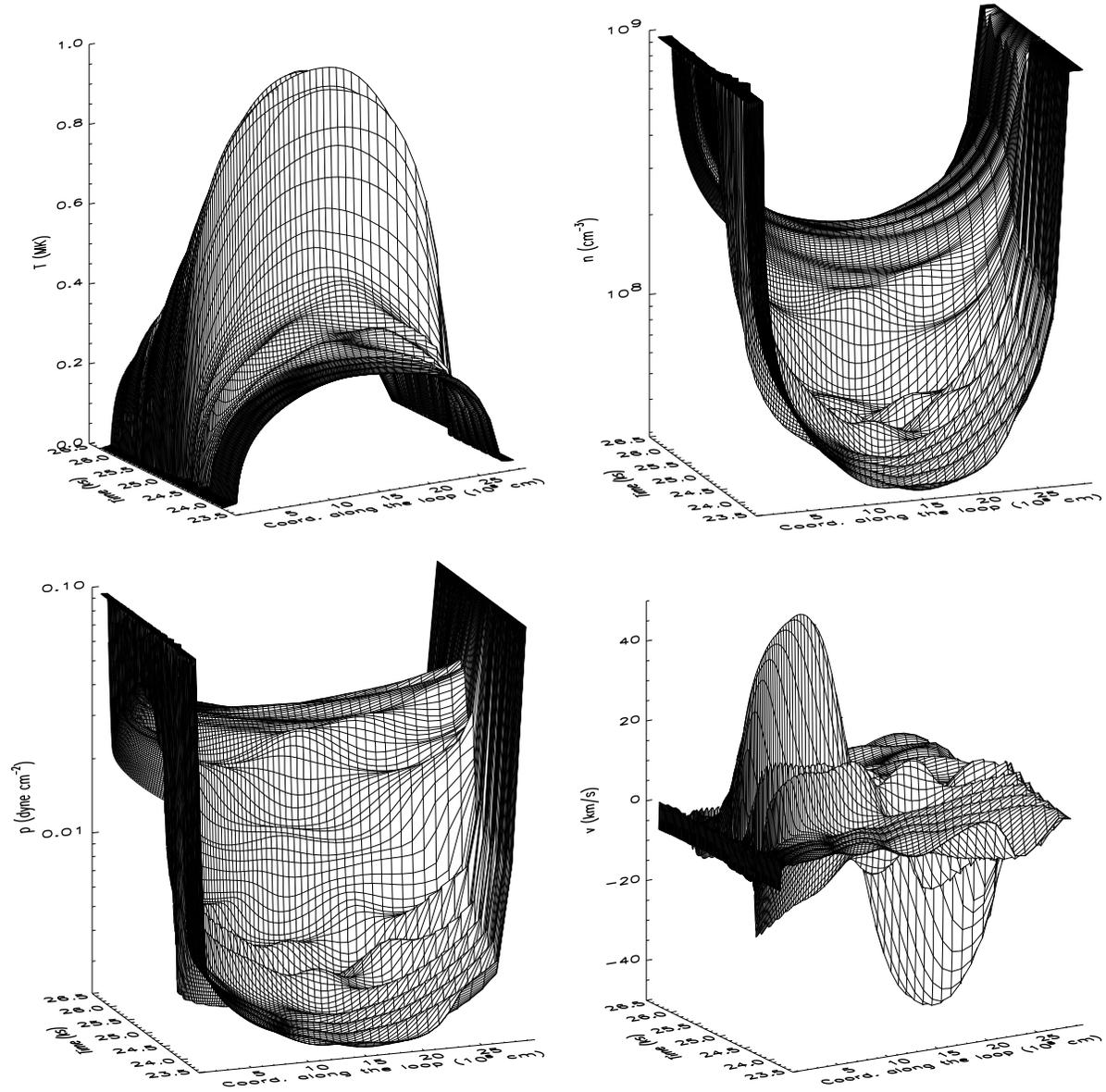,width=16cm}}
\caption[]{Evolution of the distributions of temperature, particle
density, velocity, pressure along the loop
during segment Ref1 (the fourth one in Fig.~\ref{fig:hhlc}). 
\label{fig:ref0}}
\end{figure*}

Fig.~\ref{fig:ref0} shows the evolution of the temperature, particle
density, pressure and velocity distributions along the loop obtained from
the loop simulations during segment Ref1. The temperature is steadily
below 0.2 MK until the pulse at $t \approx 24.5$ ks. Then it gradually
increases due to the enhanced heating. Fig.~\ref{fig:ref0} clearly shows
that the effects of the spatial heating structure (Fig.~\ref{fig:hh2d})
are smoothed by the efficient thermal conduction. The pulse drives also
plasma evaporation from the chromosphere, visible in the density, pressure
and velocity distributions (the negative velocity peaks indicate plasma
moving upwards from the far footpoint). The density distributions shows more
significant fluctuations traveling along the loop.

\begin{figure*}
\centerline{\psfig{figure=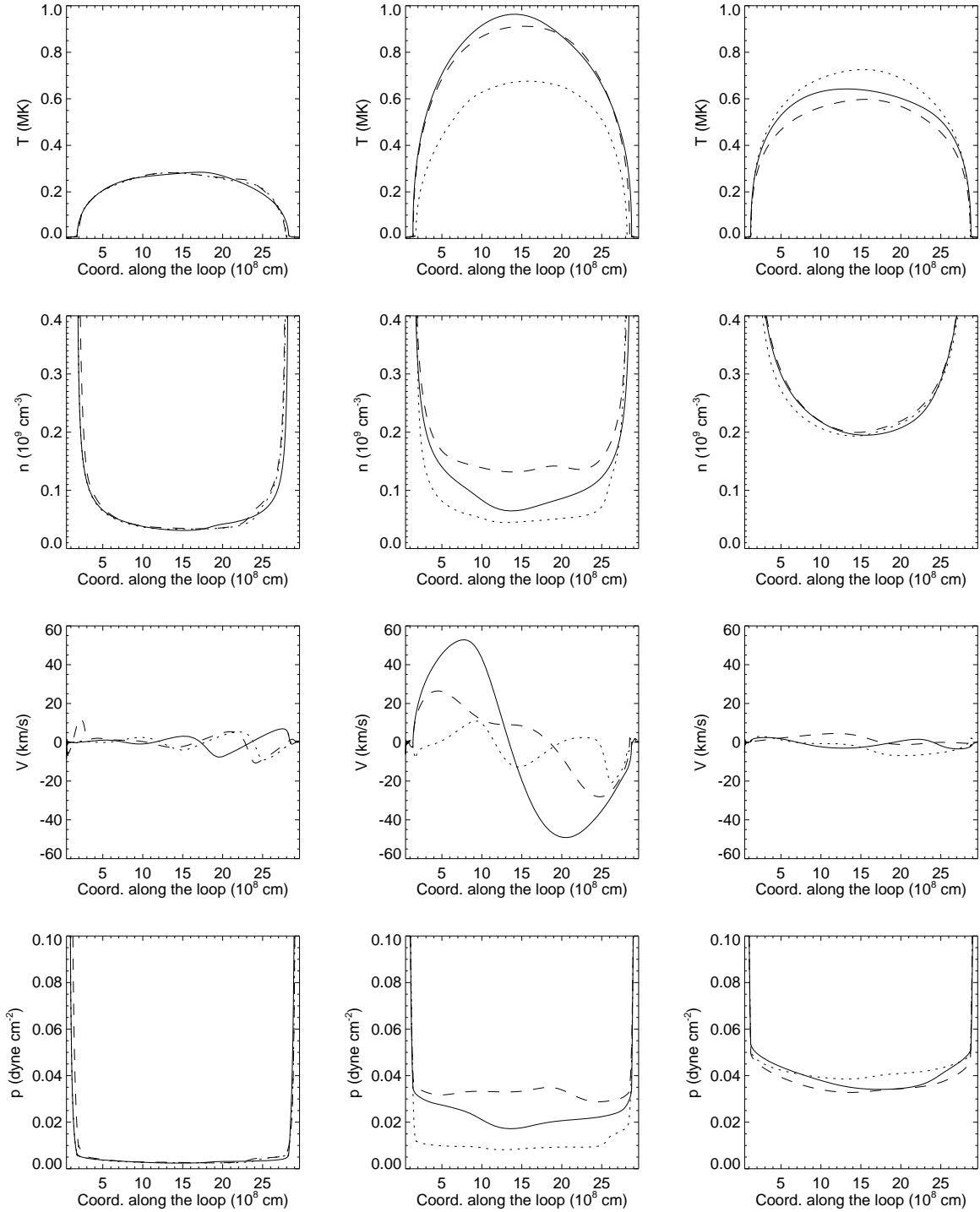,width=16cm}}
\caption[]{Distributions of temperature, particle
density, velocity, pressure along the loop
sampled during segment Ref1 (the fourth one in Fig.~\ref{fig:hhlc}) at
the three times marked 
in Fig.~\ref{fig:hh2d} (one for each column). We show the distributions at
the time ({\it solid lines}), and 100 s before ({\it dotted
lines}) and after ({\it dashed lines}).
\label{fig:ref1}}
\end{figure*}

For more quantitative information, Fig.~\ref{fig:ref1} shows selected
distributions of temperature, particle density, velocity, pressure along
the loop around the times marked in Fig.~\ref{fig:hh2d}.  Each column
of the figure shows the distributions along the loop at the exact time,
as well as 100 s before and after this time.  In the low heating state
(left column), the temperature is steadily between 0.2 and 0.3 MK
along most of the loop with a profile very similar to that of a static
loop. Also the density does not change much along the loop and is always
below $10^8$ cm$^{-3}$ in most of the loop. The distribution of plasma
velocity shows fluctuations with amplitude $\sim 10$ km/s propagating
back and forth along the loop.  During the heat pulse, the temperature
increases to about 1 MK (in $\sim 100$ s).  The distribution at the time
of the temperature maximum appears to be more peaked than in the cool
state and the position of the maximum slightly oscillates around the loop
apex. At later times ($t >25$ ks), the temperature slowly decreases and
its distribution flattens (right panel).  Asymmetric fronts of plasma
evaporation develop as the heating increases (center panel, solid line)
and the density starts to increase.  The density continues to increase
even after the temperature maximum (right panel), staying above $2 \times
10^8$ cm$^{-3}$ for a long time.  During the heat pulse, the plasma
evaporation fronts are clearly visible also in the velocity profiles:
two similar strong fronts rise from both footpoints after t=24.8 ks,
reaching a speed of about 50 km/s at intermediate positions along
the loop.  Then the plasma noticeably becomes less dynamic.  During the
heating decay, the loop slowly returns to a cool average state around 0.4
MK. The plasma velocity continues to decrease until the plasma becomes
practically static around t=25.5 ks. Then the velocity distribution gets
inverted: plasma begins to drain along the loop, at very low speed (lower
than 10 km/s).  The pressure distribution along the loop is quite stable
in the cool state. When the heating increases, the pressure increases
as well (together with the temperature and the density).  The pressure
distribution then settles to a very flat distribution during the pulse
decay at about 0.04 dyne cm$^{-2}$.

\begin{figure*}
\centerline{\psfig{figure=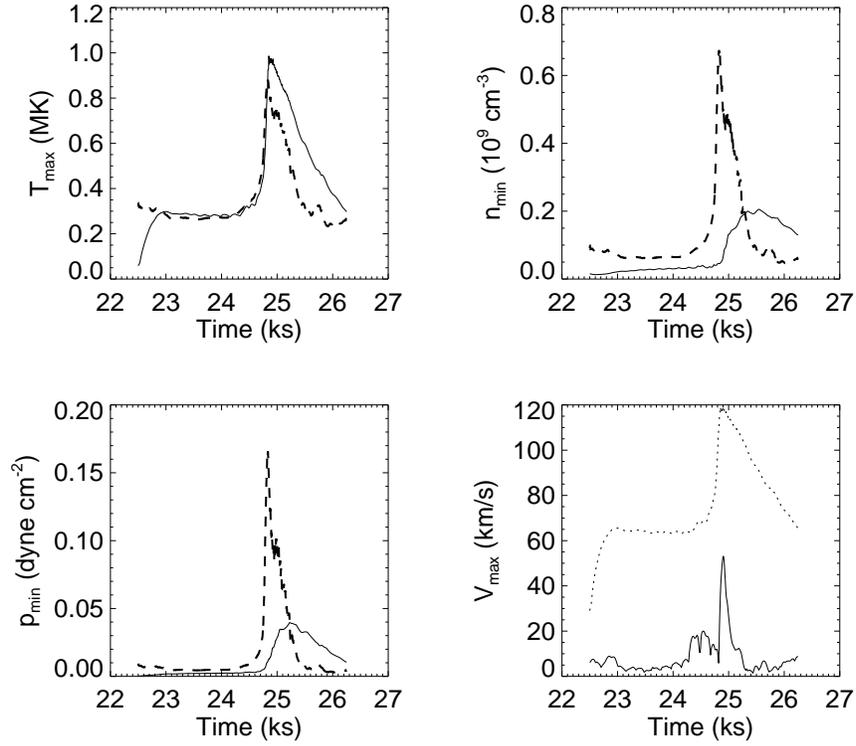,width=12cm}}
\caption[]{Evolution of the loop maximum temperature, minimum density,
minimum pressure and maximum velocity along the loop during segment Ref1. The
dashed lines
indicate the loci of the equilibrium conditions of the loop
according to the loop scaling laws and corresponding to the heating evolution
in the Fig.~\ref{fig:hhlc}. In the velocity plot, the dotted line is the
sound speed at the loop maximum temperature.
\label{fig:ref1_max}}
\end{figure*}

Fig.~\ref{fig:ref1_max} shows the evolution of the loop maximum
temperature, the loop minimum density and pressure, and of the maximum
velocity. The first three quantities are typical of the upper region of
the loop, close to the apex, the last midway between the apex and the
footpoint of the loop. The evolution of the loop maximum temperature is
globally similar to that of the average heating (Fig.~\ref{fig:hhlc}),
but much less noisy.  Consequently, it is similar also to the evolution
of the maximum temperature expected from the evolution of the average
loop heating through the loop scaling laws (Rosner et al. 1978). The
former temperature is slightly higher ($\sim 10$ \%) and decays more
slowly than the latter one. The peak temperature is different because
scaling laws assume a constant and uniform heating, while the actual
heating function in the simulation is variable and non-uniform along
the loop. The slower decay is due to the fact that the plasma response
to heating decrease is not instantaneous, and the cooling processes have
their own characteristic times.  The density enhancement due to the heat
pulse of this segment is significantly delayed ($\sim 300$ s) with respect
to the temperature increase, as typical of loop plasma evaporation. For
comparison, Fig.~\ref{fig:ref1_max} shows the equilibrium loop density
values as expected from the loop scaling laws.  The comparison clearly
shows the delay mentioned above, but emphasizes as well that during the
pulse rise the loop is significantly underdense, and becomes overdense
in the later decay phase. This is expected in dynamically heated loops:
while the heating is on, the loop is filling with plasma and therefore
below the density equilibrium conditions; when the heating stops, the loop
cools down but the plasma drains even more slowly.  The maximum pressure
has an evolution in between that of the density and of the temperature,
and explains why the plasma dynamics is time-shifted with respect to
the plasma thermal evolution.  Fig.~\ref{fig:ref1_max} shows that the
plasma velocity is constantly below 20 km/s except during the heat pulse,
when it grows to about 50 km/s. These values are well subsonic.

\begin{figure*}
\centerline{\psfig{figure=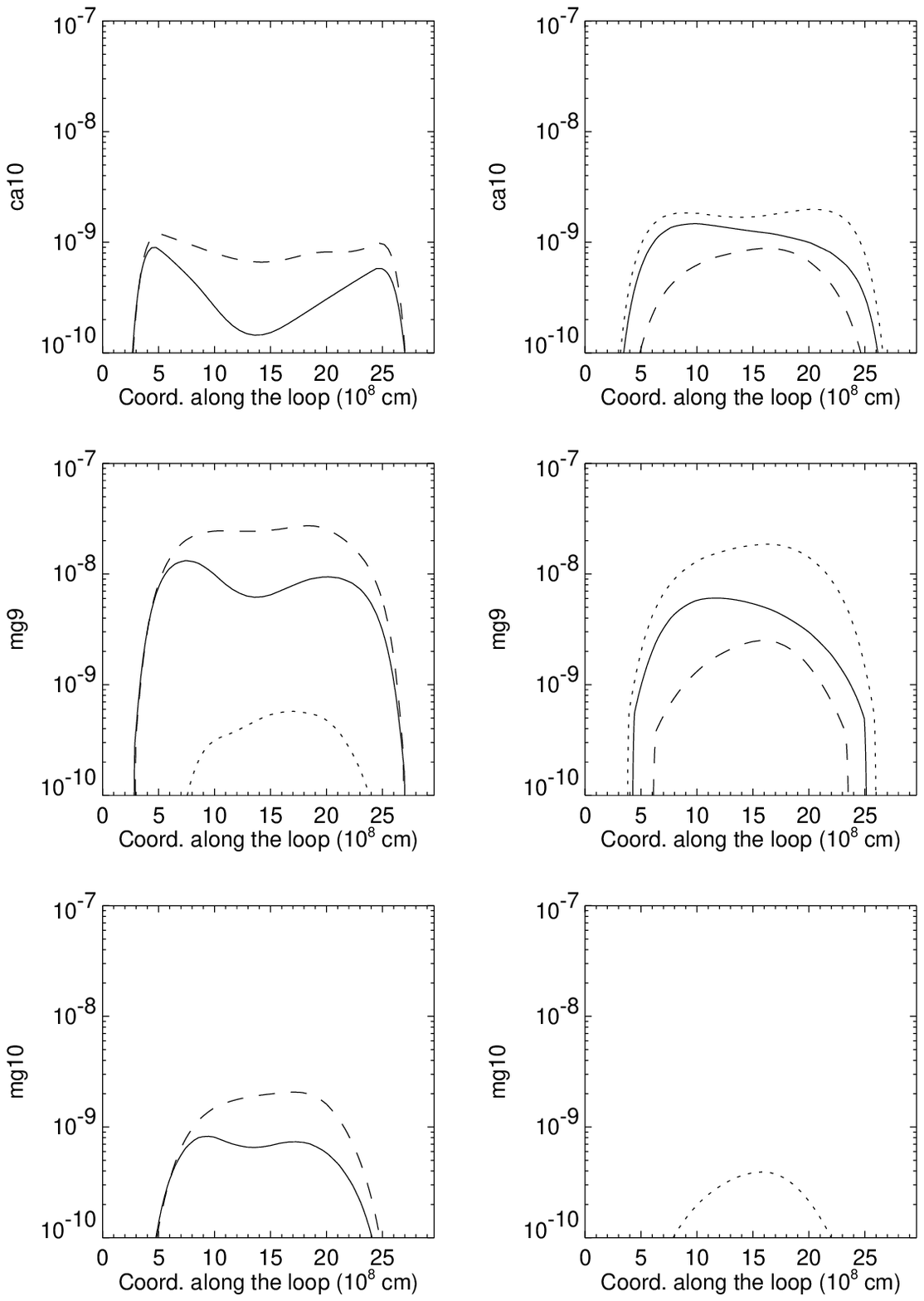,width=12cm}}
\caption[]{Emission distributions (erg cm$^{-3}$ s$^{-1}$)
along the loop in three relevant XUV lines
(Ca X 558~\AA, Mg IX 368~\AA, Mg X 625~\AA)
during segment Ref1 at the same times as the left two columns in 
Fig.~\ref{fig:ref1}. For the chosen loop parameters $10^{-10}$ erg cm$^{-3}$
s$^{-1}$ is a reasonable threshold for detection.
\label{fig:ref1_lines}}
\end{figure*}

From the output results of the hydrodynamic simulations, i.e.
distributions of temperature, density and velocity along the loop
sampled at regular time intervals, it is possible to compute the UV and
X-ray emission from the confined plasma. 
Fig.~\ref{fig:ref1_lines} shows the emission along the loop in three
representative XUV lines, i.e. Ca X 558~\AA, Mg IX 368~\AA, Mg X
625~\AA, peaking at $\log T = 5.9, 6.0$ and 6.1, respectively, at
the same times as the distributions shown in the left two columns in
Fig.~\ref{fig:ref1}. Since the line emission is sensitive both to the
temperature and to the square of the density, the emission distributions
are less uniform and fluctuate more.  This may be a distinctive signature
of this model in loop observations.  In these lines the loop is visible
for a limited time during this segment. In the hottest line (Mg X 625~\AA)
it decays very rapidly.

\subsection{High pulse}

\begin{figure*}
\centerline{\psfig{figure=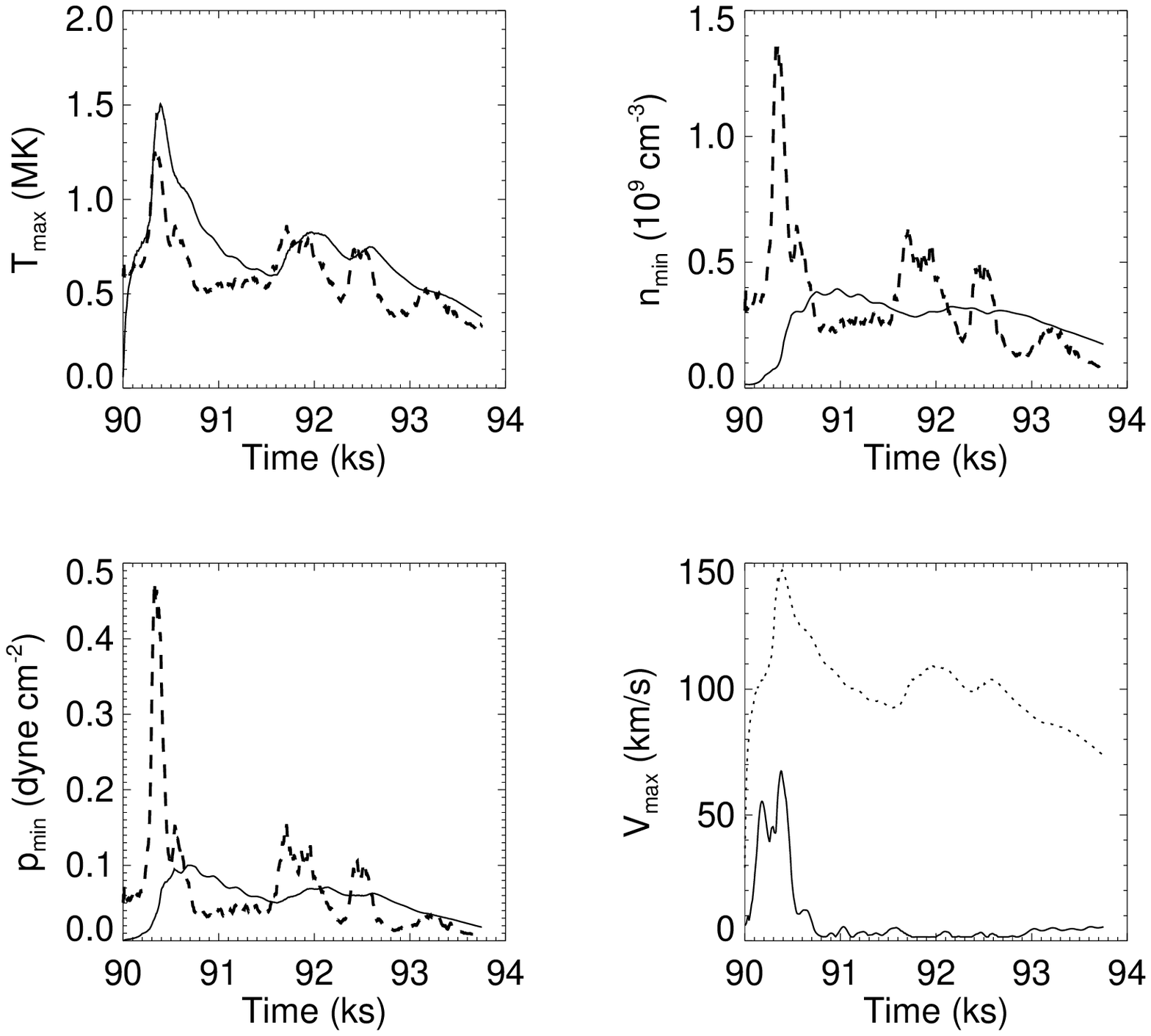,width=12cm}}
\caption[]{Evolution of the loop maximum temperature, minimum density,
minimum pressure and maximum velocity along the loop during segment RefH. The
dashed and dotted lines as in Fig.~\ref{fig:ref1_max}.
\label{fig:refh_max}}
\end{figure*}

In the course of the whole sequence of heating evolution, the most intense
heat pulse -- which we will label RefH -- occurs little after time t=90 ks
(eighth panel in Fig.~\ref{fig:hhlc}).
Fig.~\ref{fig:refh_max} shows the evolution of the loop maximum
temperature, the loop minimum density and pressure, and of the maximum
velocity, to be compared with the evolution obtained in segment Ref1 
(Fig.~\ref{fig:ref1_max}).  The loop
maximum temperature reaches 1.5 MK around time t=90.5 ks. Then it
decays below 1 MK, but stays above 0.5 MK for the rest of the segment
because of the occurrence of other minor heat pulses. The density at
the apex reaches about $4 \times 10^{8}$ cm$^{-3}$ and a pressure of
0.1 dyne cm$^{-2}$ around time t=91 ks, about 500 s later than the
temperature peak. The velocity gets above 60 km/s, always amply
subsonic.

\begin{figure*}
\centerline{\psfig{figure=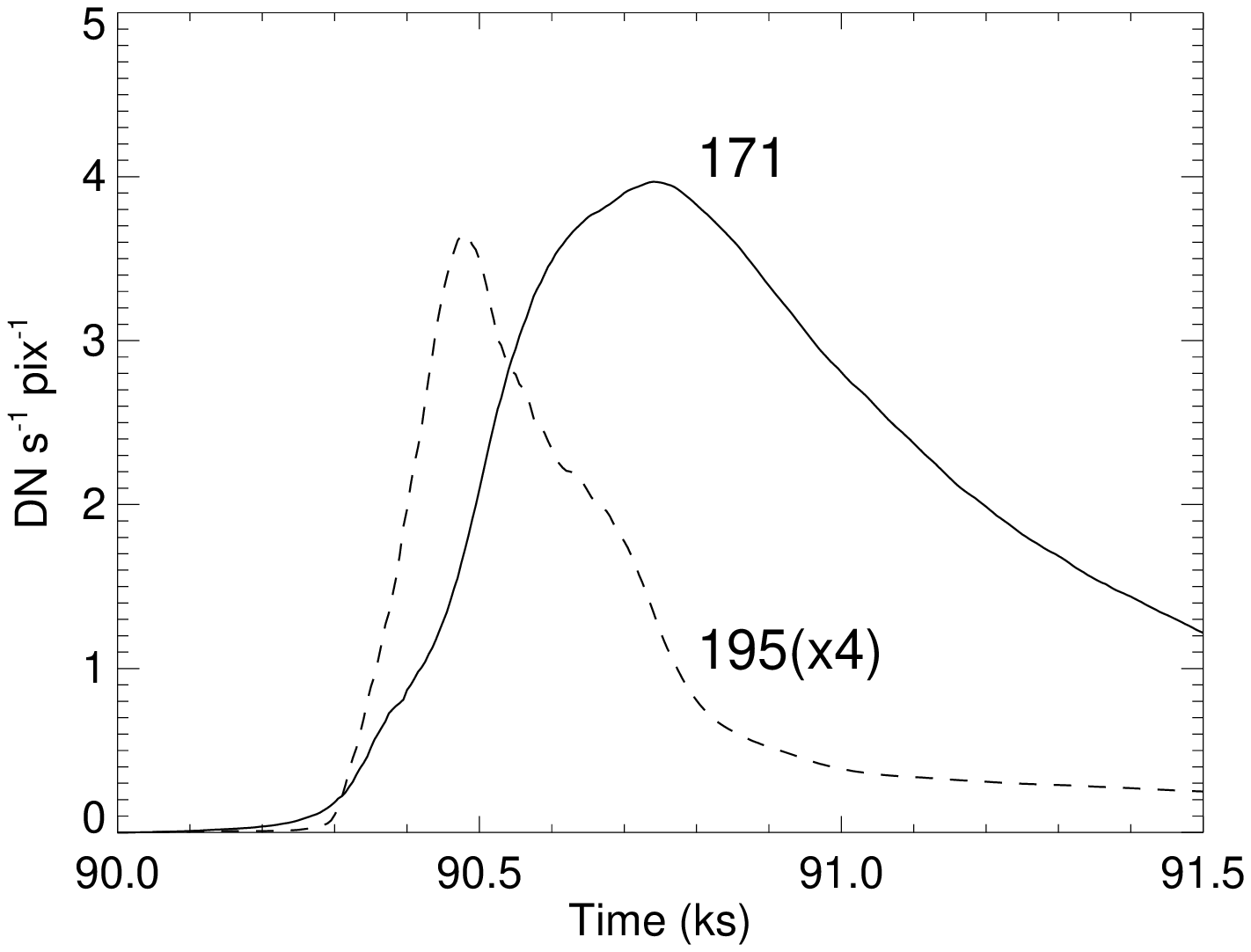,width=8cm}}
\caption[]{Light curves integrated along the whole loop during segment
RefH in the 171 A ({\it solid line}) and in the 195 ({\it dashed line})
filter bands of the TRACE telescope.  The 195 A emission is multiplied
by 4.
\label{fig:refh_lc}}
\end{figure*}

Fig.~\ref{fig:refh_lc} shows the light curves integrated along the whole
loop during segment RefH in the 171 A and in the 195 filter bands of the
Transition Region and Coronal Explorer (TRACE, Handy et al. 1999). The
light curve in the 171 A filter band resembles the evolution of the
heat pulses (although much smoother). In the 195 A filter band, only the
first pulse is significant, and only in its initial phase the emission
is significant, giving the impression of an anticipated evolution. This
evolution resembles more closely the evolution of the maximum temperature
shown in Fig.~\ref{fig:refh_max}.

\section{Discussion and conclusions}
\label{sec:discuss}

This work is devoted to exploring the effect of nanoflares due to the
magnetic energy dissipation through MHD-turbulence on the dynamic and
thermal evolution of the plasma in a coronal loop.  The parameters
considered in NMCV04, i.e. and Alfven speed of 2000 km/s corresponding
to a magnetic field of about 10 G in corona, lead to a loop with a
typical maximum temperature of $5 \times 10^5$ K. Since coronal loops are
typically observed at higher temperatures, $\geq 1$ MK, here we focus on
the effects produced by the most intense heat pulses predicted in NMCV04.
We compute in detail the hydrodynamics and thermodynamics of the loop
plasma during the pulses and analyze the results.

Although the spatial distribution of the heating has significant
fluctuations traveling along the loop and also rapid fluctuations in time,
we find that the plasma is not so fast to react and smoothes out the
fluctuations both in space and in time. We find that, under the effect
of a medium heat pulse, the loop plasma reaches $T \sim 1 MK$ and density
$\sim 0.2 \times 10^9$ cm$^{-3}$. The efficient thermal conduction makes
the plasma respond promptly to the heating deposition but also smooths
the heating fluctuations. The plasma rapidly reaches the equilibrium
temperature (according to the loop scaling laws) and then cools following
the decay of the heat pulse. The same evolution occurs for a higher heat
pulse, which produces a higher peak temperature of 1.5 MK and a higher
density of $0.5 \times 10^9$ cm$^{-3}$.  The density (and pressure) of
the plasma shows more significant fluctuations traveling along the loop
but globally responds on longer time scales. The heat pulses do not last
long enough to let the plasma reach the thermo/hydrostatic equilibrium:
the plasma is underdense during the heat pulse and overdense after the
pulse with respect to thermal equilibrium.  This density evolution
is a consequence of the impulsive heating (Winebarger et al. 2003a,
Warren et al. 2003). The speed of the plasma driven by the heat pulse
is relatively small, largely subsonic, and speeds of few tens of km/s
occur only for very few minutes.  The emission distribution in relevant
spectral lines may be relatively more sensitive to fluctuations due to
the turbulent heating and may be used to diagnose this model.  For the
highest heat pulse, our model also predicts the light curves in two
relevant TRACE filter bands to be ``out of phase" one from the other.
This phase difference is in qualitative agreement with observations
(Winebarger et al. 2003b) but also predicted by other different loop models
(Warren et al.  2003).

The heating model used here has very few free parameters (essentially
the magnetic field strength and the loop length) and depends on basic
physical effects. The shell model does not yield a detailed description of
turbulence, and cannot reproduce the energy distribution, in the direction
transverse to the magnetic field.  However, it should be adequate to
describe the behaviour of the loop integrated in the transverse direction
and the detailed energy dissipation along the loop, matching the scope
of the Palermo-Harvard loop model.

A series of questions are opened by this work. First, characterizing
features of the proposed heating are the disturbances traveling along
the loop. We have shown that observations in single spectral lines may
be sensitive to disturbances in the loop, but detecting such effects
may not be trivial with present day instruments.  Also, one may wonder
on the effect of changing the magnetic field strength; can a stronger
field lead to hotter active region loops or even major flares? Even if
the heating function may be modified with a simple scaling, this question
requires anyhow additional detailed loop modeling, since the loop plasma
evolves non-linearly under the effect of the heating, coupled with the
dynamics and the cooling processes.

As further issue to investigate, we note that the heating function
is modified by the local plasma conditions, e.g. the density
stratification and its time variation (the Alfven speed depends on
the density). Including self-consistently a feedback of the loop
plasma conditions on the energy dissipation may easily modify some
characteristics of the heating function, such as the pulse duration,
and thus influence the results.  Tackling this question requires to
couple the hybrid MHD turbulence model with the loop time-dependent
hydrodynamic model, a task planned for future work.

This first work paves the path to future works along several lines,
such as the time decomposition analysis of results and the coupling of
the heating and loop models, the comparison with observations,
encompassing the selection (or acquisition) and analysis of
observations made of long and regularly sampled image sequences.

\acknowledgements{FR and GP acknowledge support for this work from
Ministero dell'Istruzione, Universit\`a e
Ricerca.  G.N., F.M. and P.V. acknowledge partial support by the MIUR
(Ministero dell Istruzione, dell'Universit\`a e della Ricerca) through
a National Project Fund (cofin 2002) and by the European Community
within the Research Training Network  Turbulence in Space Plasmas,
Theory, Observation and Simulation. RMHD numerical calculations were
performed in the framework of HPCC (Center for High Performance
Computing) of the University of Calabria.  }

\end{document}